\documentclass[
 reprint,
superscriptaddress,
 amsmath,amssymb,
 aps,
 prl,
floatfix,
]{revtex4}
\usepackage{float}
\usepackage{fancyhdr}
\usepackage{setspace}
\usepackage{graphicx}
\usepackage[svgnames]{xcolor}
\usepackage{dcolumn}
\usepackage{bm}
\usepackage{graphicx}
\begin{document}
\pagestyle{fancy}
\fancyhf{}
\fancyfoot[R]{\thepage}
\preprint{APS/123-QED}
\title{Departure from the Wiedemann-Franz Law in WP$_2$ Driven by Mismatch in T-square Resistivity Prefactors}

\author{Alexandre Jaoui}
\email{alexandre.jaoui@espci.fr}
\affiliation{JEIP, USR 3573 CNRS, Coll\`ege de France,
PSL Research University, 11, Place Marcelin Berthelot,
75231 Paris Cedex 05, France}
\affiliation{Laboratoire de Physique et Etude des Mat\'eriaux (CNRS/UPMC), Ecole Sup\'erieure de Physique et de Chimie Industrielles, 10 Rue Vauquelin, 75005 Paris, France}

\author{Beno\^it Fauqu\'e}
\affiliation{JEIP, USR 3573 CNRS, Coll\`ege de France,
PSL Research University, 11, Place Marcelin Berthelot,
75231 Paris Cedex 05, France}
\affiliation{Laboratoire de Physique et Etude des Mat\'eriaux (CNRS/UPMC), Ecole Sup\'erieure de Physique et de Chimie Industrielles, 10 Rue Vauquelin, 75005 Paris, France}

\author{Carl Willem Rischau}
\affiliation{Laboratoire de Physique et Etude des Mat\'eriaux (CNRS/UPMC), Ecole Sup\'erieure de Physique et de Chimie Industrielles, 10 Rue Vauquelin, 75005 Paris, France}
\affiliation{Department of Quantum Matter Physics (DQMP), University of Geneva, 24 Quai Ernest-Ansermet, 1211 Geneva 4, Switzerland}

\author{Alaska Subedi}
\affiliation{Centre de Physique Th\'eorique, Ecole Polytechnique, CNRS, Universit\'e Paris-Saclay, F-91128 Palaiseau, France}
\affiliation{Coll\`ege de France, 11 Place Marcelin Berthelot, 75005 Paris, France}

\author{Chenguang Fu}
\affiliation{Max Planck Institute for Chemical Physics of Solids, N\"othnitzer Str. 40, 01187 Dresden, Germany}

\author{Johannes Gooth}
\affiliation{Max Planck Institute for Chemical Physics of Solids, N\"othnitzer Str. 40, 01187 Dresden, Germany}

\author{Nitesh Kumar}
\affiliation{Max Planck Institute for Chemical Physics of Solids, N\"othnitzer Str. 40, 01187 Dresden, Germany}

\author{Vicky S\"{u}{\ss}}
\affiliation{Max Planck Institute for Chemical Physics of Solids, N\"othnitzer Str. 40, 01187 Dresden, Germany}

\author{Dmitrii L. Maslov}
\affiliation{Department of Physics, University of Florida, Gainesville, Florida 32611, USA}

\author{Claudia Felser}
\affiliation{Max Planck Institute for Chemical Physics of Solids, N\"othnitzer Str. 40, 01187 Dresden, Germany}

\author{Kamran Behnia}
\email{kamran.behnia@espci.fr}
\affiliation{Laboratoire de Physique et Etude des Mat\'eriaux (CNRS/UPMC), Ecole Sup\'erieure de Physique et de Chimie Industrielles, 10 Rue Vauquelin, 75005 Paris, France}
\affiliation{II. Physikalisches Institut, Universit\"at zu K\"oln, 50937 K\"oln, Germany}

\date{\today}

\begin{abstract}
The Wiedemann-Franz (WF) law establishes a link between heat and charge transport due to electrons in solids. The extent of its validity in presence of inelastic scattering is a question raised in different contexts. We report on a study of the electrical, $\sigma$, and thermal, $\kappa$, conductivities in WP$_2$ single crystals. The WF holds at 2 K, but a downward deviation rapidly emerges upon warming. At 13 K, there is an exceptionally large mismatch between Lorenz number and the Sommerfeld value. We show that this is driven by a fivefold discrepancy between the $T$-square prefactors of electrical and thermal resistivities, both caused by electron-electron scattering. This implies the existence of abundant small-scattering-angle collisions between electrons, due to strong screening. By quantifying the relative frequency of collisions conserving momentum flux, but degrading heat flux, we identify a narrow temperature window where the hierarchy of scattering times may correspond to the hydrodynamic regime.
\end{abstract}

\maketitle

\section{Introduction}
The electrical conductivity of a metal $\sigma$ and its thermal counterpart $\kappa$ are linked to each other by the Wiedemann-Franz (WF) law, provided that the heat carried by phonons is negligible and electrons do not suffer inelastic scattering. This law states that the ratio of the two conductivities divided by temperature should be equal to a universal number set by fundamental constants. The validity of the WF law is expected both at very low temperatures, where elastic scattering by disorder dominates, and above the Debye temperature, where scattering by phonons becomes effectively elastic. At intermediate temperatures, inelastic scattering is known to degrade thermal current more efficiently than the electrical current \cite{ziman}. Experiments have found a zero-temperature validity combined to a downward departure in elemental metals (due to electron-phonon scattering) \cite{PhysRev.119.1869,doi:10.1063/1.4997034} as well as in correlated metals (because of electron-electron scattering) \cite{PhysRevLett.94.216602,PhysRevLett.73.3294}. During the past decade, the search for a possible breakdown of the WF law near a quantum critical point \cite{Pfau2012} motivated high-resolution experiments, which verified its zero-temperature validity within experimental margin and quantified the deviation at finite temperature \cite{Seyfarth2008,Machida2013,Reid2014,Taupin2015}.

Gooth \emph{et al.}$\,$\cite{gooth} have recently reported on thermal transport in micrometric samples of WP$_2$ down to 5K and found a drastic breakdown of the WF law. WP$_2$ is a type-II Weyl semimetal with a room-temperature Residual Resistivity Ratio (RRR) expressed in five digits and an impressively large magnetoresistance \cite{WP2}. The observation raised fundamental questions regarding the relevance of the scattering-based theory of charge and entropy transport by mobile electrons to this non-trivial solid. The possible link between WF breakdown and electron hydrodynamics is a subject of attention \cite{galitski,lucas,principi,Lucas2018,Coulter2018}.

In this paper, we present a study of thermal conductivity in bulk millimetric single crystals of WP$_2$. By performing concomitant measurements of thermal and electrical transport between 2K and 40K, we find that: (i) The WF law is obeyed at 2K, but a drastic downward deviation of exceptional amplitude emerges at higher temperatures; (ii) Thanks to the low-temperature data, one can distinguish between the contributions to the thermal and electrical resistivities arising from electron-electron and electron-phonon scattering;  iii) The downward deviation arises because of a large (fivefold) difference between the amplitudes of the T-square prefactors in the two (electrical and thermal) resistivities due to electron-electron scattering. We conclude that electron-electron scattering is the origin of the exceptionally large downward deviation from the Wiedemann-Franz law. This can  happen if small-angle momentum-relaxing scattering events are unusually frequent. Thus, the semi-classical transport theory is able to explain a large mismatch between Lorenz number and Sommerfeld number at finite temperature. However, the large T-square thermal resistivity caused  by momentum-conserving scattering among electrons, together with the long mean-free-path of the electrons, opens a window for entering into the hydrodynamic regime.

\color{black} 

\section{Results}
\begin{figure}[H]
\makebox{\includegraphics[width=0.9\textwidth]
    {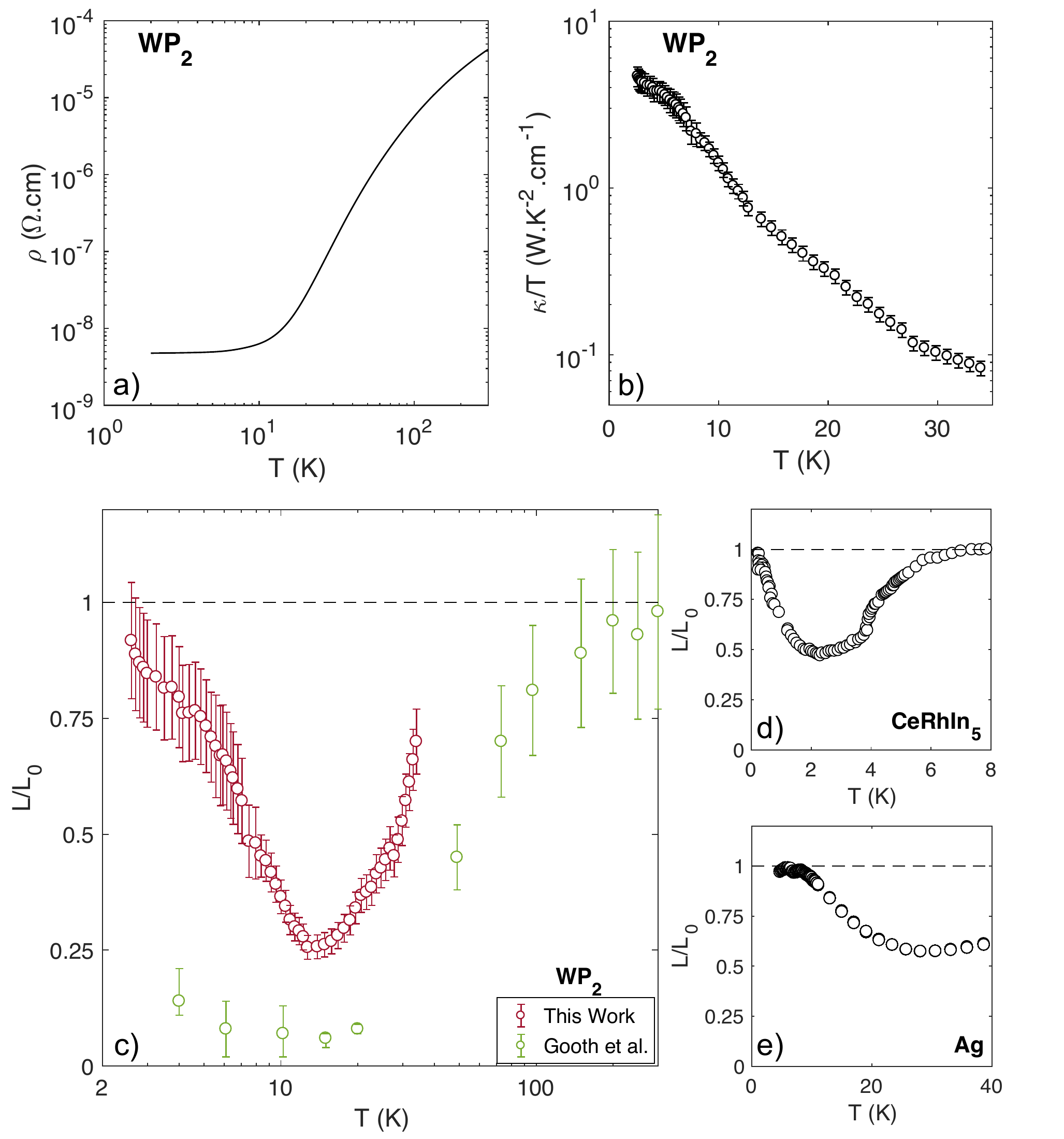}}
    \caption{a) Resistivity of WP$_2$, $\rho$, measured along the \emph{a}-axis, as a function of temperature. b) Thermal conductivity divided by temperature, $\frac{\kappa}{T}$ ($j_Q$//{\textit a}-axis) as a function of temperature in the same sample with the same electrodes. c) Ratio of Lorenz, $L(T)=\frac{\kappa}{T\sigma}$ to Sommerfeld $L_0=2.44\times10^{-8}$ $W.\Omega.K^{-2}$, numbers as a function of temperature (red). The data reported for a micro-ribbon of WP$_2$ \cite{gooth} are shown in green. d) $L(T)/L_0$ as a function of temperature in CeRhIn$_5$ \cite{PhysRevLett.94.216602}. e) $L(T)/L_0$ as a function of temperature for an Ag wire with a 50 $\mu$m diameter and 99.99\% purity. WP$_2$ and Ag were measured with the same experimental setup. Error bars are due to the error on the measurement of the resistance of the thermometers which are used to evaluate the temperature gradient in the sample.}
    \label{Fig.1}
\end{figure}

Fig.\ref{Fig.1}.a shows the resistivity as a function of temperature in a WP$_2$ single crystal. The RRR for this sample is $\rho(300K)/\rho(2K)=9600$. The residual resistivity $\rho_0$ of the different samples was found to lie between 4 and 6 n$\Omega$.cm. With a carrier density of $2.5\times 10^{21}$ cm$^{-3}$ \cite{gooth}, this implies a mean-free-path in the range of 70 to 140 $\mu$m, and, given the dimensions of the sample, a proximity to the ballistic limit.

The temperature dependence of $\kappa/T$, the thermal conductivity divided by temperature, is plotted in panel (b) of the Figure \ref{Fig.1}. Note that in our whole temperature range of study, the phonon contribution to heat transport is negligible (See the Supplemental Material). The extracted Lorenz number, $L(T)=\frac{\kappa\rho}{T}$, is to be compared with the Sommerfeld number $L_0=\frac{\pi^2}{3}(\frac{k_{B}}{e})^2$. As seen in Fig.\ref{Fig.1}.c, according to our data, $L/L_0$ is close to 0.5 at 40K and decreases with decreasing temperature until it becomes as low as 0.25 at 13K, in qualitative agreement with the observation originally reported by Gooth \emph{et al.} \cite{gooth}, who first reported on a very low magnitude of the $L/L_0$ ratio in WP$_2$. As seen in the Figure \ref{Fig.1}.c, however, the two sets of data diverge at low temperature and we recover the expected equality between $L$ and $L_0$ at low temperature.

Comparison with two other metals, Ag and CeRhIn$_5$, is instructive. Fig.\ref{Fig.1}.d displays the temperature dependence of L/L$_0$ in the heavy fermion antiferromagnet, CeRhIn$_5$ as reported by Paglione \textit{et al. }\cite{PhysRevLett.94.216602}. The $L/L_0$ ratio, close to unity at 8K, decreases with decreasing temperature and becomes as low as 0.5 at 2K, before shooting upwards and attaining unity around 100mK. In Ag, as seen in Fig.\ref{Fig.1}.e, which presents our data obtained on a silver wire, a similar downward deviation of the L/L$_0$ ratio is detectable. Close to unity below 8K, it decreases with warming and attains a minimum of 0.6 at 30K before increasing again.

It is also instructive to recall the case of semi-metallic bismuth, in which thermal transport is dominated by phonons. In such a compensated system, an ambipolar contribution to the thermal conductivity, arising from a counter-flow of heat-carrying electrons and holes, was expected to be present \cite{UHER,Korenblit}. An ambipolar diffusion would have led to an upward deviation of L/L$_0$ from unity. However, Uher and Goldsmid \cite{UHER} found (after subtracting the lattice contribution) that L/L$_0<1$ in bismuth, which indicates that there is no ambipolar contribution to the thermal conductivity. The absence of a significant phononic contribution in our data makes the interpretation even more straightforward, and we also find no evidence for ambipolar heat transport in WP$_2$. The reason is that the electron and hole gases are degenerate both in Bi (below room temperature) and WP$_2$ (for all temperatures of interest), and thus the ambipolar contribution is small in proportion to $T/E_F$.

\begin{figure}[H]
\centering
\makebox{\includegraphics[width=0.35\textwidth]
    {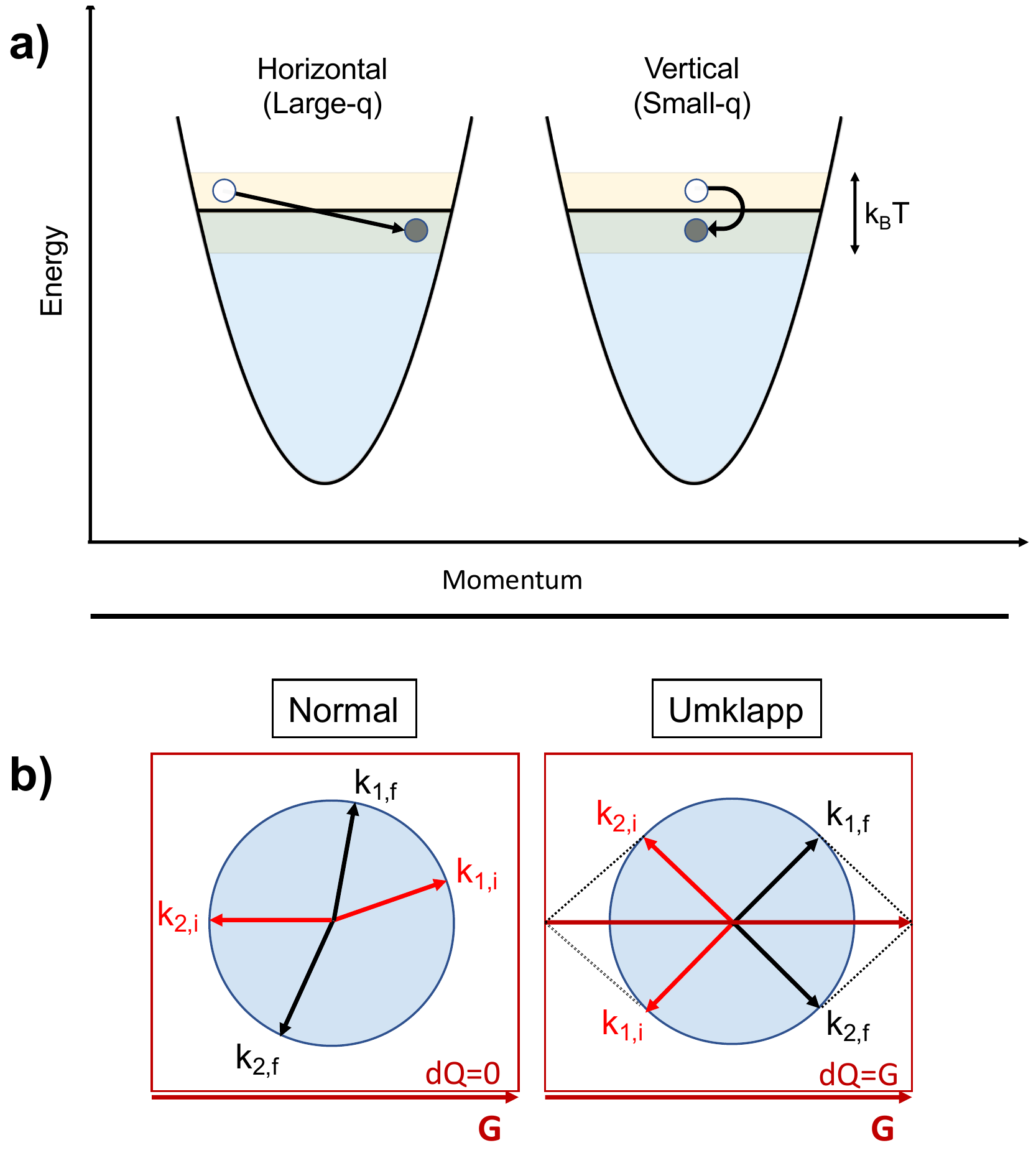}}
    \caption{a) A schematic representation highlighting the difference between horizontal and vertical inelastic scattering. Both kind of processes degrade heat transport. Their effect on momentum transport is however very different. Arrows indicate scattering from one state to another. b) A sketch of Normal and  Umklapp scattering in the case of a simple circular Fermi surface (blue), inside a rectangular Brillouin zone (red). They differ by the balance of quasi-momentum exchange after collision. $\vec{dQ}=\vec{k_{2f}}+\vec{k_{1f}}-\vec{k_{2i}}-\vec{k_{1i}}$ is zero in a normal process and equal to a unit vector of the reciprocal lattice in an Umklapp one.}
     \label{Fig.2N}
\end{figure}

The scattering-based Boltzmann picture provides an explanation for such downward deviations. Thermal and electrical transport are affected in different ways by inelastic collisions labeled as ``horizontal'' and ``vertical'' (See Fig.\ref{Fig.2N}.a). In a horizontal scattering event, the change in the energy of the scattered carrier is accompanied by a drastic change of its momentum. Such a large-q process degrades both charge and heat currents. A vertical process, on the other hand, is a small-q scattering event, which marginally affects the carrier momentum, but modifies its energy as strongly as a horizontal process of similar intensity. In the case of momentum transport, the presence of a  (1-$\cos\theta$) pondering factor disfavors small angle scattering. No such term exists for energy transport. This unequal importance of vertical events for electrical and thermal conductivities, pulls down the $L(T)/L_0$ ratio at finite temperature and generates a finite-temperature breakdown of the Wiedemann-Franz law \cite{ziman}. Such a behavior was observed in high-purity Cu half a century ago \cite{PhysRev.119.1869}, in other elements, such as Al and Zn \cite{doi:10.1063/1.4997034}, in heavy-fermion metals such as UPt$_3$ \cite{PhysRevLett.73.3294}, CeRhIn$_5$ \cite{PhysRevLett.94.216602} or CeCoIn$_5$ \cite{Seyfarth2008} as well as in magnetically-ordered elements like Ni \cite{nickel} or Co \cite{cobalt}.

On the microscopic level, two distinct types of vertical scattering have been identified. The first is electron-phonon scattering \cite{ziman}, relevant in elemental metals. At low-temperatures, the Bloch-Gr\"{u}neisen picture of electron-phonon scattering yields a $T^5$ electric resistivity and a $T^3$ thermal resistivity. The higher exponent for charge transport is due to the variation of the typical wave-vector of the thermally-excited phonons with temperature: $q_{ph}=\frac{k_BT}{\hbar v_s}$. Small-angle phonon-scattering becomes more frequent with cooling. Therefore, phonons' capacity to degrade a momentum current declines faster than their ability to impede energy transport. This power-law difference leads to $L(T)/L_0 < 1$ in the intermediate temperature window (below the Debye temperature), when phonon scattering dominates over impurity scattering, but all phonons are not thermally excited. A second source of q-selectivity concerns momentum relaxing electron-electron scattering (See Fig.\ref{Fig.2N}.b). The quadratic temperature dependence of resistivity in a Fermi liquid is a manifestation of such scattering \cite{Pal}. This is because the phase space for collision between two fermionic quasi-particles scales with the square of temperature. Since the total momentum before and after collision is conserved, electron-electron collisions degrade the flow of momentum only when the scattering is accompanied by losing part of the total momentum to the lattice. Two known ways for such a  momentum transfer are often invoked \cite{Lin}. The first is Baber mechanism, in which electrons exchanging momentum belong to two distinct reservoirs and have different masses. The second is an Umklapp process, where the change in the momentum of the colliding electrons is accompanied by the loss of one reciprocal lattice wave-vector (Fig.\ref{Fig.2N}.b). Abundant small-angle electron-electron scattering (which could be either Umklapp or Baber-like)  would generate a mismatch in prefactors of the $T$-square resistivities with the electrical prefactor lower than the thermal one. This is a second route towards $L(T)/L_0 < 1$,  prominent in correlated metals \cite{PhysRevLett.19.167,PhysRev.185.968}.

In order to determine what set of microscopic collisions causes the downward deviation from the WF law in WP$_2$, we identified and quantified various contributions to the thermal and electrical resistivities of the system.

Fig.\ref{Fig.2} shows the electrical resistivity, $\rho$, and thermal resistivity, $WT$, as a function of $T^2$. In order to keep the two resistivities in the same units and comparable to each other, we define $WT= \frac{T L_0}{\kappa}$, as in reference \cite{PhysRevLett.94.216602}. One can see that at low temperatures, the temperature-induced increase in $\rho$ and $WT$ is linear in $T^2$, confirming the presence of a $T$-square component in both quantities. The intercept is equal in both plots, which means that the WF law is valid in the zero-temperature limit. But the two slopes are different and the deviation from the low-temperature quadratic behavior occurs at different temperatures and in different fashions.

\section{Discussion}

Admitting three distinct contributions (scattering by defects, electrons and phonons) to the electrical and thermal resistivities, the expressions for $\rho$ and $WT$ become:

\begin{equation}
\rho\, = \,\rho_0\,+\,A_2T^2\,+\,A_5T^5
\end{equation}
\begin{equation}
WT\, = \,W_0T\,+\,B_2T^2\,+\,B_3T^3
\end{equation}

\begin{figure}[H]
\centering
\makebox{\includegraphics[width=0.4\textwidth]
    {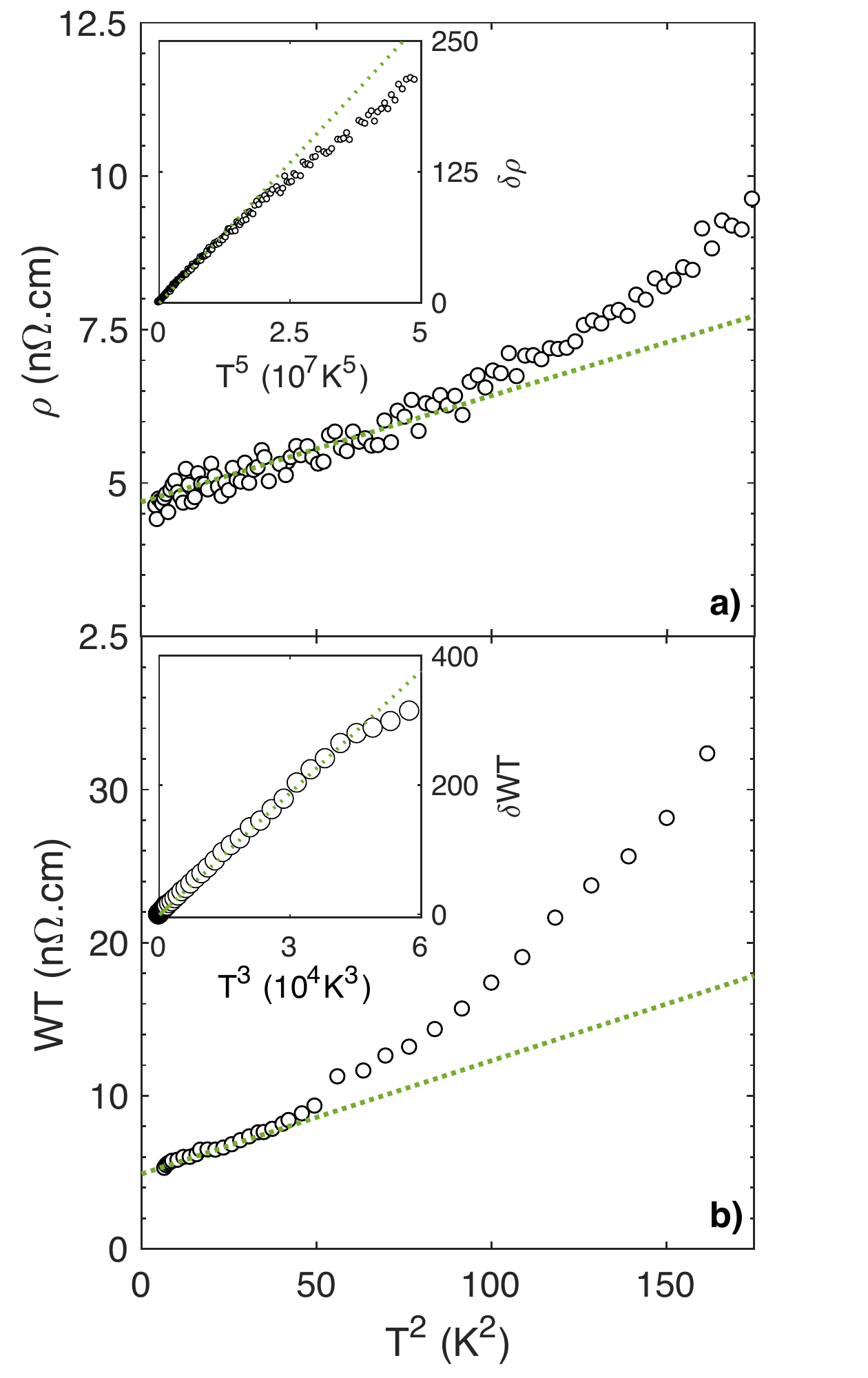}}
    \caption{a) Electrical resistivity, $\rho$, as a function of $T^2$. The dotted line is a linear fit to the $T<10$K data. The inset shows $ \delta\rho\,=\,\rho\,-\,\rho_0\,-\,AT^2$ as a function of $T^5$ expressed in n$\Omega$.cm. b) Thermal resistivity, $WT=\frac{L_0T}{\kappa}$, as a function of $T^2$. The dotted line is a linear fit to the $T<8$K data. The inset shows $\delta WT\,=\,WT\,-\,W_0T\,-\,BT^2$ as a function of $T^3$ expressed in n$\Omega$.cm.}
     \label{Fig.2}
\end{figure}

\begin{figure}[H]
\centering
\makebox{\includegraphics[width=0.70\textwidth]
    {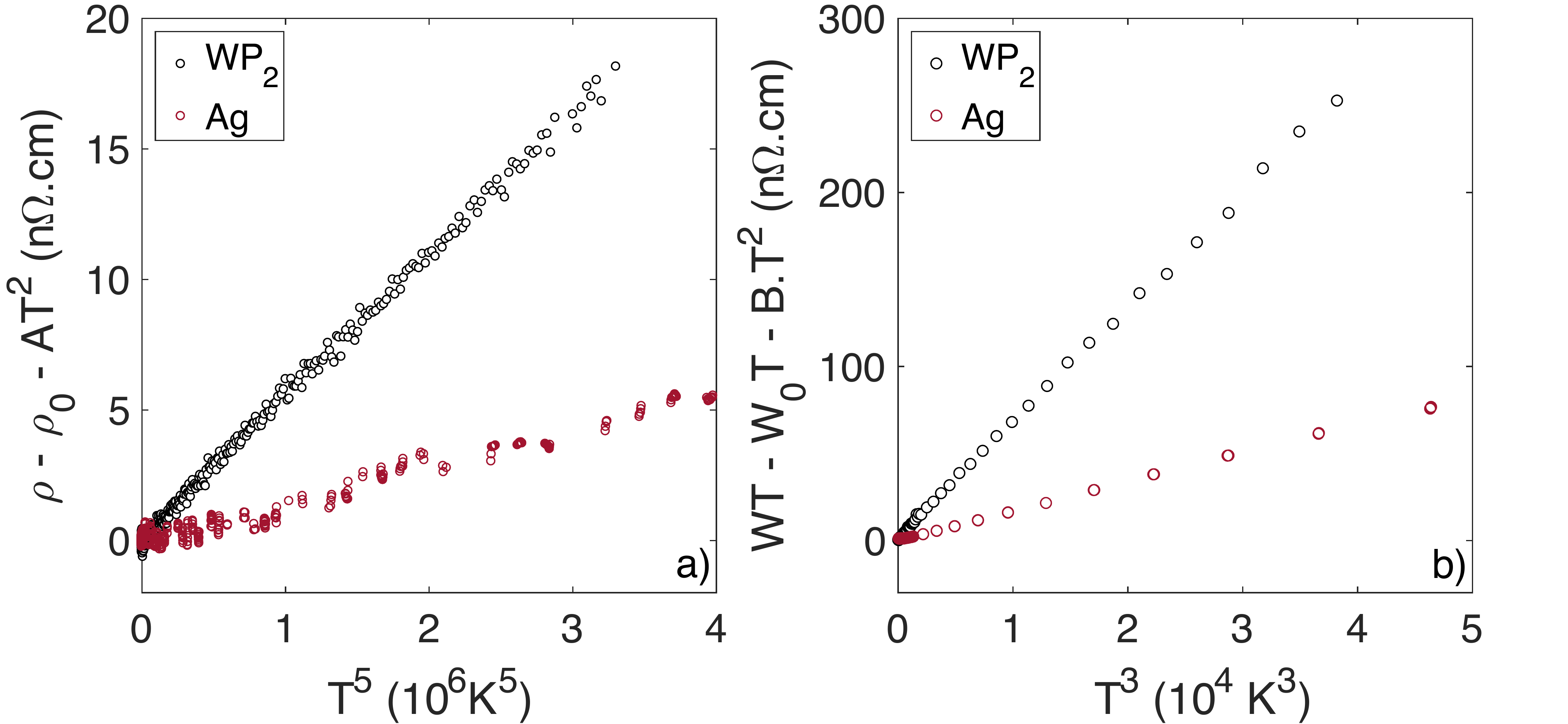}}
    \caption{a) Inelastic electrical resistivity $\delta\rho$ caused by phonon scattering in WP$_2$ (Black) and Ag (Red) as a function of $T^5$. b) Inelastic thermal resistivity, $\delta WT$, in WP$_2$ (Black) and Ag (Red) as a function of $T^3$. The ratio of the two slopes is similar for heat and charge transport. }
     \label{Fig.3}
\end{figure}

We assume these scattering mechanisms to be additive. Note that since the data are limited to a temperature window in which $T> \frac{\hbar}{k_B \tau}$, no Altshuler-Aronov corrections are expected \cite{PhysRevB.64.214204}. As seen above, $\rho_0 = W_0T$, but $A_2 \neq B_2$. The insets in Fig.\ref{Fig.2} show that $\rho-\rho_0-A_2T^2$ is linear in $T^5$ and $WT-W_0T\,-\,B_2T^2$ is proportional to $T^3$, in agreement with what is expected from equations (1) and (2).
\\It is now instructive to compare WP$_2$ and Ag to examine the possible role played by inelastic phonon scattering. Fig.\ref{Fig.3} compares the amplitude of the $T^5$ terms in WP$_2$ and Ag. As seen in the Figure \ref{Fig.3}, the amplitude of both $A_5$ and $B_3$ is larger in WP$_2$. More quantitatively, $A_5$(WP$_2$)/$A_5$(Ag)$\,=\,$3.4 and $B_3$(WP$_2$)/$B_3$(Ag)$\,=\,$3.6. In other words, the $B_3$ and $A_5$ ratios of WP$_2$ and Ag are similar in magnitude, which implies that phonon scattering is not the origin of the unusually low magnitude of the Lorenz number in WP$_2$.

Having ruled out a major role played by phonon scattering in setting the low magnitude of $L/L_0$, let us turn our attention to electron-electron scattering. As stated above, the prefactors of the T-square terms in $\rho$ and $WT$, namely $A_2$ and $B_2$, are unequal. The ratio $A_2$/$B_2$ is as low as 0.22,  well below what was observed in other metals, such as CeRhIn$_5$ ($A_2$/$B_2 \simeq0.4$) \cite{PhysRevLett.94.216602}, UPt$_3$ ($A_2$/$B_2 \simeq0.65$) \cite{PhysRevLett.73.3294}, or nickel ($A_2$/$B_2 \simeq0.4$) \cite{nickel}. This feature, which pulls down the magnitude of the $L/L_0$ ratio in WP$_2$, may be due to unusually abundant vertical events (involving a small change in the wave-vector of one of the colliding electrons), which could be either Umklapp or inter-band involving collisions between hole-like and electron-like carriers belonging to different pockets.

\begin{figure}[H]
\centering
\makebox{\includegraphics[width=0.65\textwidth]
    {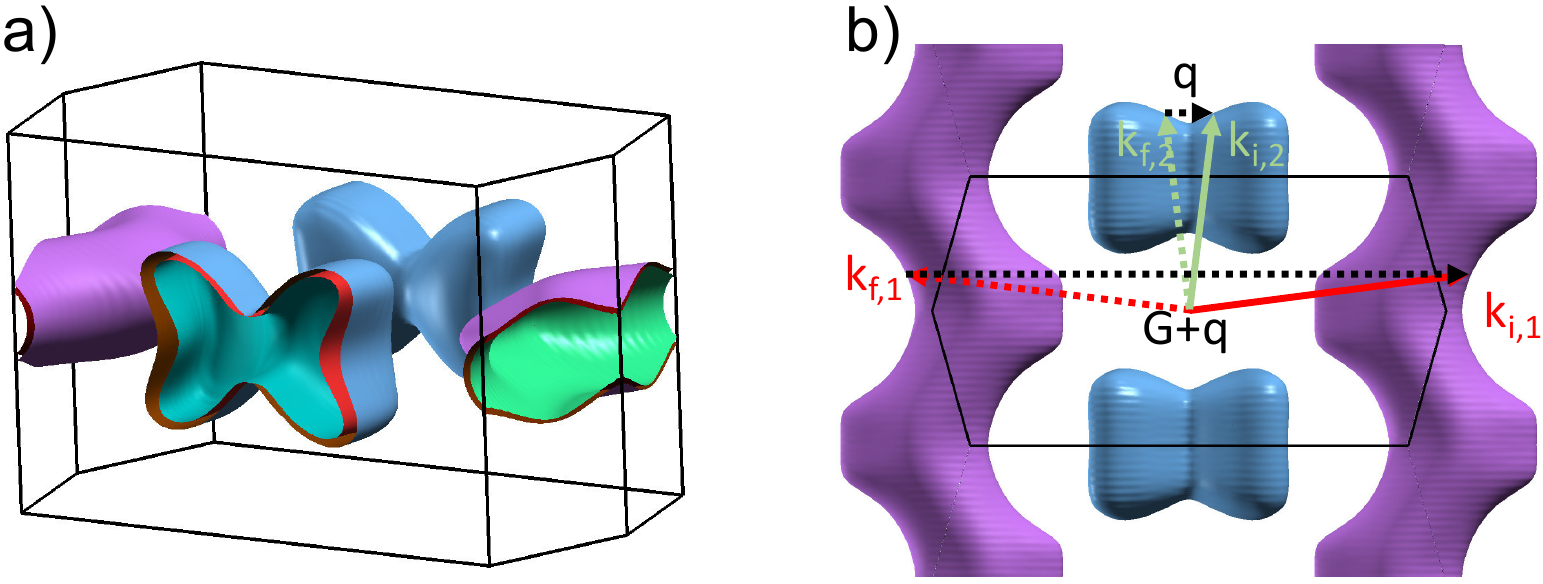}}
		\caption{a) DFT-computed 3D Fermi Surface of WP$_2$. b) Top view of the Fermi surface. Arrows illustrate wave-vectors during an inter-band Umklapp and small-angle scattering event. $\vec{k_{i,n}}$ and $\vec{k_{f,n}}$ are carrier momenta, $\vec{k_{f,2}}-\vec{k_{i,2}}=\vec{q}$ and $\vec{k_{f,1}}-\vec{k_{i,1}}=\vec{G}+\vec{q}$.}
		\label{Fig.4}
\end{figure}

To have electron-electron collisions which are simultaneously small-angle, Umklapp and intra-band, one needs a Fermi surface component located at the zone boundary \cite{maslov}. Interestingly, as seen in  Fig.\ref{Fig.4}, this is the case of WP$_{2}$. The figure shows the Fermi surface obtained by our DFT calculations (See supplement), consistent with previous reports \cite{damascelli,balicas}. It is composed of 2 hole-like and 2 electron-like pockets, each located at the boundary of the Brillouin zone. Such a configuration allows abundant intra-band low-q Umklapp scattering.
According to previous theoretical calculations \cite{PhysRevLett.19.167, PhysRev.185.968}, the weight of small-angle scattering can pull down the $A_2$/$B_2$ (and  the L/L$_0$) ratio. However, the lowest number found by these theories ($\simeq $0.38) is well above what was found here by our experiment on  WP$_2$ ($A_2$/$B_2 \simeq0.22$), as  well as what was reported long ago in the case of tungsten \cite{PhysRevB.3.3141} (See the supplement).

Following the present experimental observation, Li and Maslov showed \cite{maslov_unpublished} that in a compensated metal with a long-range Coulomb interaction among the charge carriers, the Lorenz ratio is given by 
\begin{equation}
L/L_0 = (\kappa/k_F)^2/2
\end{equation}
where $\kappa$ is the (inverse) screening length and $k_F$ is the (common) Fermi momentum of the electron and hole pockets. By assumption, $\kappa \ll k_F$ and thus $L/L_0$ can be arbitrarily small in this model.

\begin{figure}[H]
\centering
\makebox{\includegraphics[width=0.62\textwidth]
    {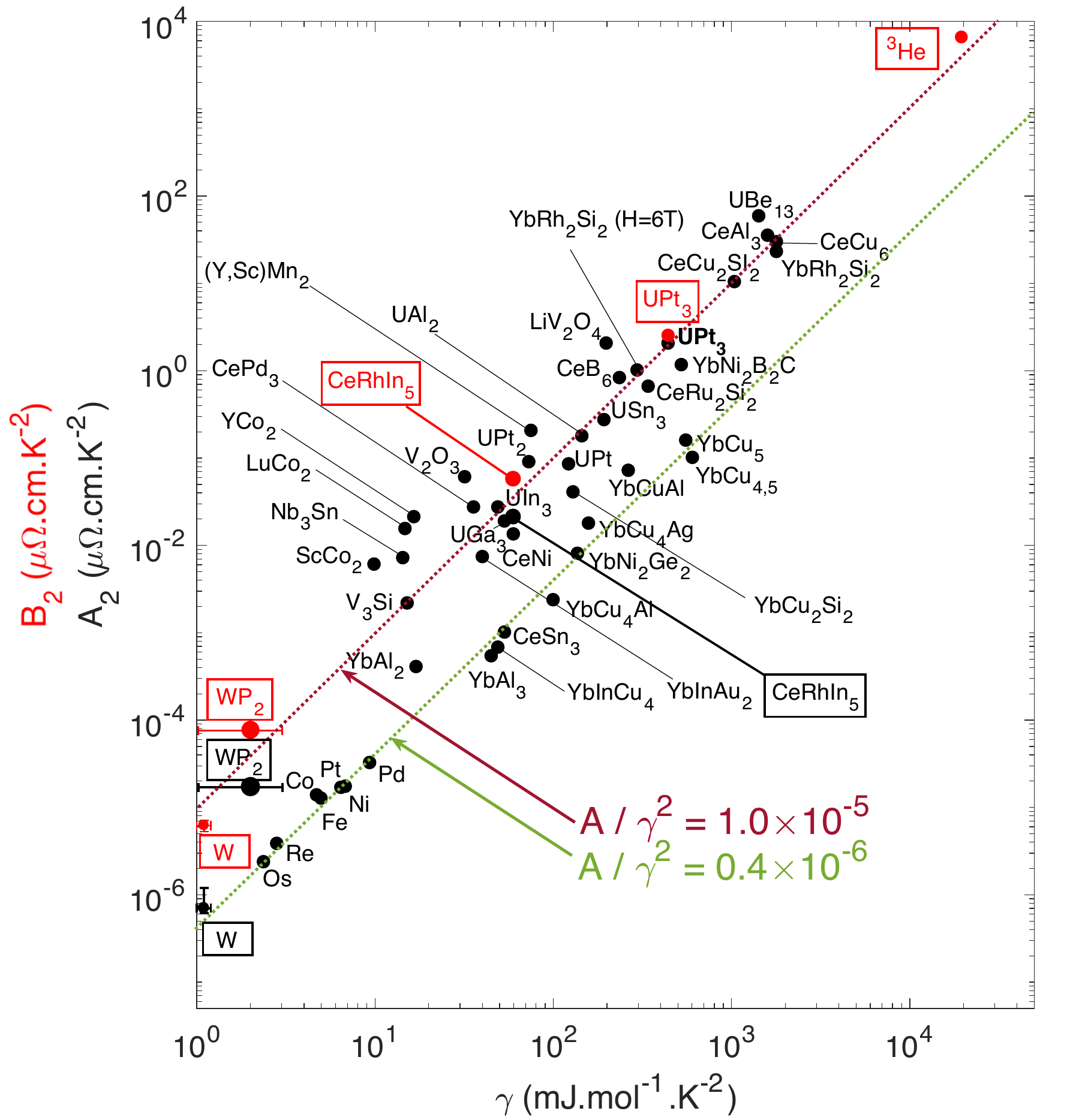}}
		\caption{a) The Kadowaki-Woods plot with A$_2$ and B$_2$ plotted as a function of fermionic specific heat $\gamma$( C =$\gamma$T). To the plot compiled by  Tsujii \emph{et al.} \cite{tsujii}, we have added the data for $^3$He \cite{PhysRevB.29.4933,PhysRevLett.18.737,wheatley:jpa-00213853}, WP$_2$ (This work), W \cite{PhysRevB.3.3141} and CeRhIn$_5$ \cite{PhysRevLett.94.216602,PhysRevB.65.224509}. These compounds are indicated by boxes. }
		\label{Fig.5}
\end{figure}

Let us now turn our attention to the possibility that WP$_2$ enters the hydrodynamic regime \cite{gooth}. In order to address this question, let us first  recall what is known in the case of normal-liquid $^3$He. The latter presents a thermal conductivity inversely proportional to temperature \cite{PhysRevB.29.4933} (strictly equivalent to our $WT$ being proportional to $T^2$) and a viscosity proportional to $T^{-2}$ \cite{Black} at very low temperatures. Both features are caused by fermion-fermion collisions \cite{PhysRevB.35.8425}, which are normal and conserve total momentum. As one can see in Fig.\ref{Fig.5}, the magnitude of B$_2$ (prefactor of the thermal T-square resistivity) in $^3$He, in CeRhIn$_5$, in WP$_2$ and in W plotted \textit{vs. $\gamma$}, the fermionic specific heat, lies close to the universal Kadowaki-Woods plot. This means that while A$_2$ quantifies the size of  momentum-relaxing collisions and B$_2$ is a measure of energy-relaxing, yet \textit{momentum-conserving} collisions, both scale roughly with the size of the phase space for fermion-fermion scattering, which (provided a constant fermion density) is set by $\gamma^2$ . As a consequence, the magnitude of B$_2$ opens a new window to determine where one may expect electron hydrodynamics.

The hydrodynamic regime \cite{PhysRevLett.106.256804,Moll} of electronic transport (identified long ago by Gurzhi \cite{gurzhi}) requires a specific hierarchy of scattering times. Momentum-conserving collisions should be more frequent than boundary scattering and the latter more abundant than momentum-relaxing collisions. Let us show that this hierarchy can be satisfied in our system thanks to the combination of an unusually low $A_2/B_2$ ratio and low disorder. The combination of a residual resistivity as low as 4 n$\Omega$.cm and a carrier density of $2.5\times10^{21}$ cm$^{-3}$  according to \cite{gooth} (compared to $2.9\times 10^{21}$ cm$^{-3}$ according to our DFT calculations) implies that we are at the onset of the ballistic limit. It yields a mean-free-path of 140 $\mu$m. This is to be compared to the sample width and thickness of 0.1 mm. 

Like in many other cases \cite{Shoenberg}, the Dingle temperature of quantum oscillations yields a mean-free-path much shorter than this. A particularly large discrepancy between the Dingle and transport mobilities has been observed in low-density semi-metals such as Sb \cite{Fauque_Sb}. In the system under study, the difference is as large as three orders of magnitude \cite{WP2}. This is presumably because of a very long screening length, weakening large-angle scattering and helping momentum conservation along long distances. 

\begin{figure}[H]
\centering
\makebox{\includegraphics[width=0.49\textwidth]
    {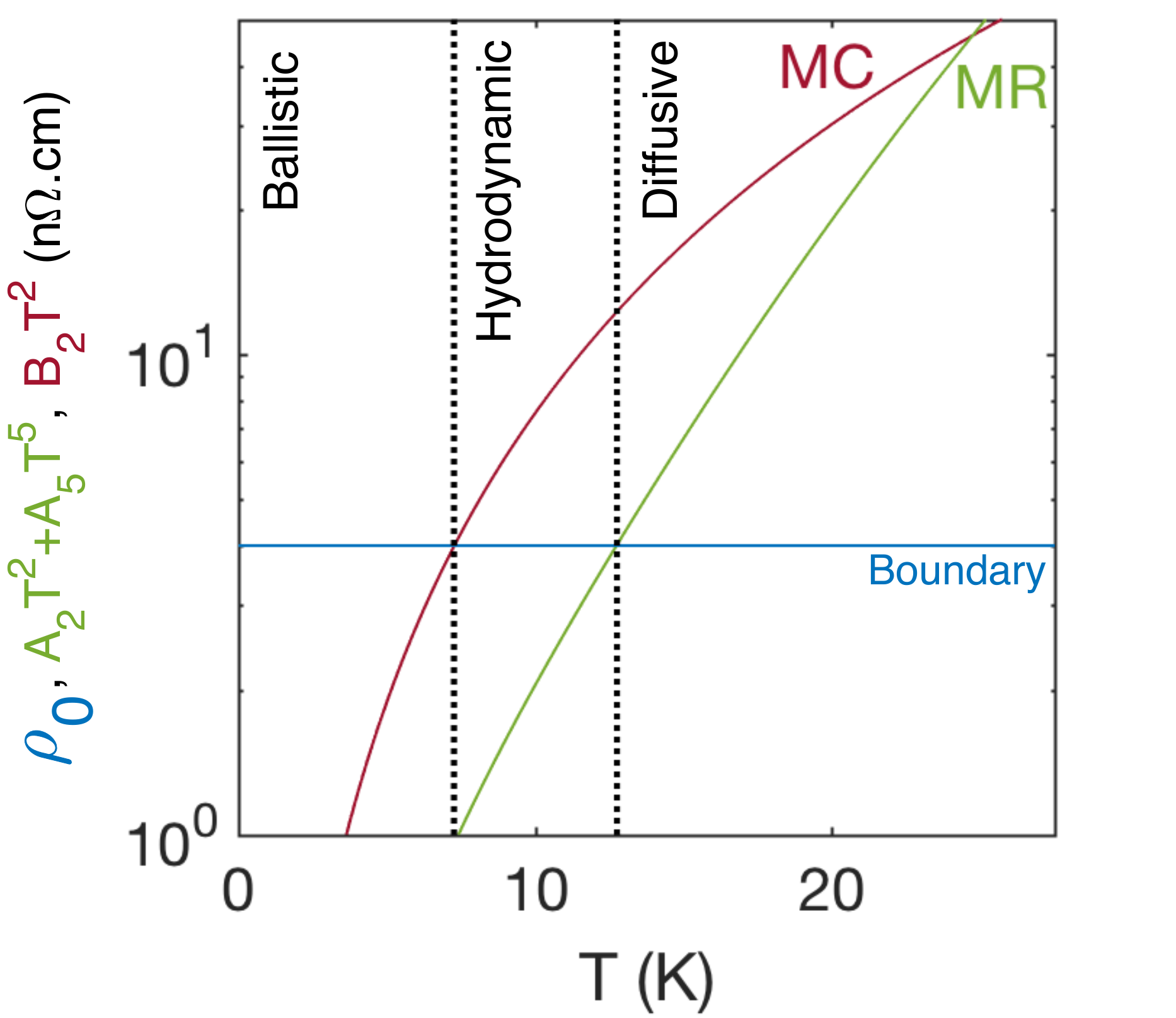}}
		\caption{The magnitude of B$_{2}T^{2}$, proportional to momentum-conserving (MC) electron-electron collisions is compared to $\rho_{imp}$+A$_{2}T^{2}+$A$_{5}T^{5}$, which is proportional to momentum-relaxing (MR) collisions by electrons and phonons and $\rho_{0}$, which is a measure of boundary scattering. In a limited temperature window (7$<$T$<$13 K), the hierarchy for hydrodynamic regime is satisfied. }
		\label{Fig.6}
\end{figure}

This feature, combined with the fact that momentum-conserving collisions are 4-5 times more frequent than momentum relaxing ones, implies that the system satisfies the required hierarchy of scattering times in a limited temperature window, as one can seen in Fig.\ref{Fig.6}. This figure compares the temperature dependence of momentum-relaxing collisions (with other electrons and phonons), momentum-conserving collisions (among electrons) and the boundary scattering. The three terms are represented by their contributions to resistivity, convertible to scattering rates by the same material-dependent factor. Note the narrowness of the temperature window and the modesty of the difference between the three scattering rates. Note also that the hydrodynamic regime coincides with the observed minimum in L/L$_0$ representing an excess of momentum flow in comparison to energy flow. In the hydrodynamic scenario, this coincidence is not an accident. However, the position and the width of this window are not solidly set. Assuming that the residual resistivity is not entirely fixed by the boundary scattering (i.e $\rho_{0}=\rho_{00}+\rho_{imp}$) would shift this temperature window and beyond a threshold $\rho_{imp}$, the window will close up. 

In purely hydrodynamic transport, momentum relaxation occurs only at the boundary of the system. Momentum-conserving collisions then set the magnitude of the viscosity and the fluid drifts in presence of an external force. However, this does not happen in WP$_2$ or in any other metal, because the finite $B_2/A_2$ ratio means that momentum-relaxing events are not absent. In our hydrodynamic regime, an electron traveling from one end of the sample to the other suffers few collisions and four-out-of-five of them conserve momentum. Because the three scattering times (momentum-conserving, momentum-relaxing and boundary) are of the same order of magnitude, any hydrodynamic signature would lead to modest corrections to what can be described in the diffusive or ballistic regimes, such as subtle departures in size-dependent transport properties \cite{Moll}. 

One message of this study is that a finite-temperature departure from the WF law by itself cannot be a signature of hydrodynamic transport, but thermal transport can be used to quantify the relative weight of momentum-conserving collisions and to identify where to expect eventual hydrodynamic features. Specifically, our study highlights two features which were not explicitly considered in previous discussions about the hydrodynamics of electrons. First, as for phonons \cite{BECK}, the hydrodynamic regime for electrons is expected to occur in a finite temperature window squeezed between the ballistic and diffusive regimes. Second, the phase spaces for momentum-relaxing and momentum-conserving collisions for electrons follow the same (T-square) temperature dependence. This is in a contrast to the case of phonons where Umklapp scattering vanishes exponentially with temperature whereas normal scattering follows a power law \cite{ACKERMAN}. This difference makes electron hydrodynamics more elusive in comparison with its phononic counterpart \cite{MACHIDA}.

We note also that the two solids showing anomalously low $L/L_0$ (W and WP$_2$) are those in which the $T=0$ ballistic limit is accessible and a hydrodynamic window can open up. Future studies on samples with different dimensions \cite{gooth} using a four-contact measurement setup are necessary to reach a definite conclusion.

In summary, we found that WP$_2$ obeys the Wiedemann-Franz law at 2K, but there is a large downward deviation which emerges at higher temperatures. We recalled that the dichotomy between charge and heat transport is ubiquitous in metallic systems, since low-q scattering affects heat conduction more drastically than charge transport. The exceptionally low magnitude of $L/L_0$ ratio mirrors the discrepancy between the amplitude of T-square prefactors in thermal and electrical resistivities. The large difference between momentum-conserving and momentum-relaxing collisions among electrons opens a narrow temperature window where the hierarchy of scattering times conforms to hydrodynamic requirements.

\section{Methods}
The samples used in this study were needle-like single-crystals (grown along the {\textit a}-axis). Their typical dimensions were $1-2 \times 0.1 \times 0.1$ mm$^3$. The samples are similar to those detailed in \cite{WP2} : they were grown by chemical vapor transport. Starting materials were red phosphorous (Alfa-Aesar, 99.999\%) and tungsten trioxide (Alfa-Aesar, 99.998\%) with iodine as a transport agent. The materials were taken in an evacuated fused silica ampoule. The transport reaction was carried out in a two-zone-furnace with a temperature gradient of 1000$^{\circ}C$ (T$_1$) to 900$^{\circ}C$ (T$_2$) for several weeks. After reaction, the ampoule was removed from the furnace and quenched in water. The metallic needle-like crystals were later characterized by X-ray diffraction.
The measurements were performed with a standard one-heater-two-thermometers set-up, with Cernox chips, allowing to measure thermal conductivity $\kappa$ and the electrical resistivity $\rho$ with the same electrodes and the same geometrical factor. Contacts were made with $25\,\mu$m Pt wires connected via silver paste with a contact resistance ranging from 1 to 10 $\Omega$. The electric and heat currents were injected along the {\textit a}-axis of the sample. By studying three different samples with different RRRs, we checked the reproducibility of our results (see the supplemental material).
\\Supplementary information : Supplementary information accompanies the paper on the npj Quantum Materials
website.
\vspace{1\baselineskip}

\section{Acknowledgements}
We are indebted to Bernard Castaing and Jacques Flouquet for stimulating discussions. 
\section{Competing Interests}
Conflict of interest : The authors declare no conflict of interest.

\section{Authors' contributions}
AJ carried out the thermal and electrical conductivity measurements. WR, AJ and BF built the probe. VS made the sample. CF and NK performed specific heat measurements. CF, JG and KB initiated the collaboration. AS carried out the DFT calculations. AJ, BF and KB analyzed the data with input from DM. All authors participated in the discussions leading to the paper, which was written by KB and AJ.

\section{Funding}
This project was funded by Fonds-ESPCI and supported by a grant from R\'egion Ile-de-France. K.B. acknowledges support by the National Science Foundation under Grant No. NSF PHY17-48958. B.F acknowledges support from Jeunes Equipes de l'Institut de Physique du Coll\`ege de France (JEIP). D.L.M. acknowledges support from NSF DMR-1720816.

\section{Data Availability}
All data supporting the findings of this study are available from the corresponding authors A.J. and K.B. upon request.

\bibliography{bibliography.bib}

\pagebreak
\section{Figures}

\vspace{5\baselineskip}
\vspace{5\baselineskip}

\pagebreak
\section{Supplemental Material for 'Departure from the Wiedemann-Franz Law in WP$_2$ Driven by Mismatch in T-square Resistivity Prefactors'}

\subsection{Samples Measured}
We measured 3 different single crystals of WP$_2$, all grown in the same batch.
The sample presented in the corpus is S3. In table \ref{table1} we present their geometries and basic electrical characteristics. 
\\All the samples were measured along the same direction, with $j\parallel({\textit a}$-axis) for both heat and electrical currents.
The result for the electrical conductivity in all three samples are shown in Fig.\ref{fig1}. Besides the residual resistivity $\rho_0$, we see that both the $T^2$-dependent and $T^5$-dependent terms are equal from one sample to another. 
\\From Fig.\ref{fig2} we can deduce that the $T^2$-dependent and $T^3$-dependent terms of the thermal resistivity, $WT$, are also equivalent in the three different samples. We thus confirm that besides the residual terms, the electrical and thermal resistivities are reproducible from one sample to another with comparable size.

\begin{table}[h!]
\begin{tabular}{|c||c|c|c|c|c|}
 \hline
 Sample & Length (mm)& Width ($\mu$m) & Thickness ($\mu$m) & $\rho_0$ (n$\Omega$.cm) & RRR =$\rho(300K)/\rho(2K)$\\ [0.5ex] 
 \hline\hline
 S1 & 1.5 & 80-100 & 110 & 3.94 &11200\\
 \hline
 S2 & 1.9 & 90-100 &110 & 5.85 &7600\\
 \hline
 S3 & 0.9& 80-110 &120 & 4.69 & 9600\\
 \hline
\end{tabular}
\caption{Presentation of the different WP$_2$ samples}
\label{table1}
\end{table}

\begin{center}
\begin{figure}[H]
\makebox{\includegraphics[width=0.90\textwidth]
    {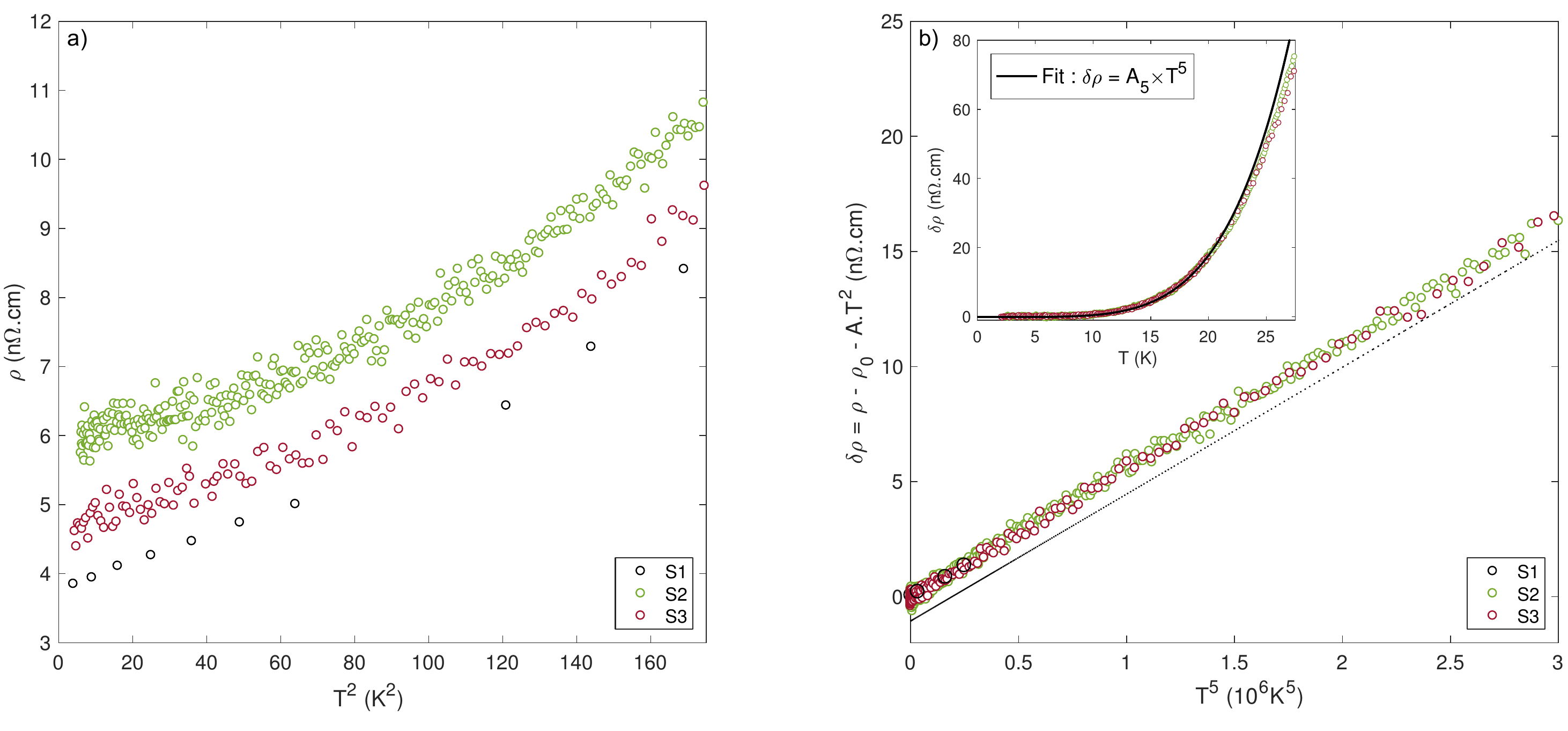}}
    \caption{a) Resistivity $\rho$, measured along the \emph{a}-axis, of WP$_2$ as a function of $T^{2}$ for the three samples S1, S2 and S3. b) Phonon contribution to the electrical resistivity $\delta\rho\,=\,\rho\,-\,\rho_0\,-\,AT^2$ as a function of $T^5$ for the same three samples. Inset shows a fit of $\delta\rho$ to a $T^5$ law with $A_5=3.9\times10^{-15}$ $\Omega$.cm.K$^{-5}$. We observe a downward deviation for $T>20K$.}
    \label{fig1}
\end{figure}
\onecolumngrid
\begin{figure}[H]
\makebox{\includegraphics[width=0.97\textwidth]
    {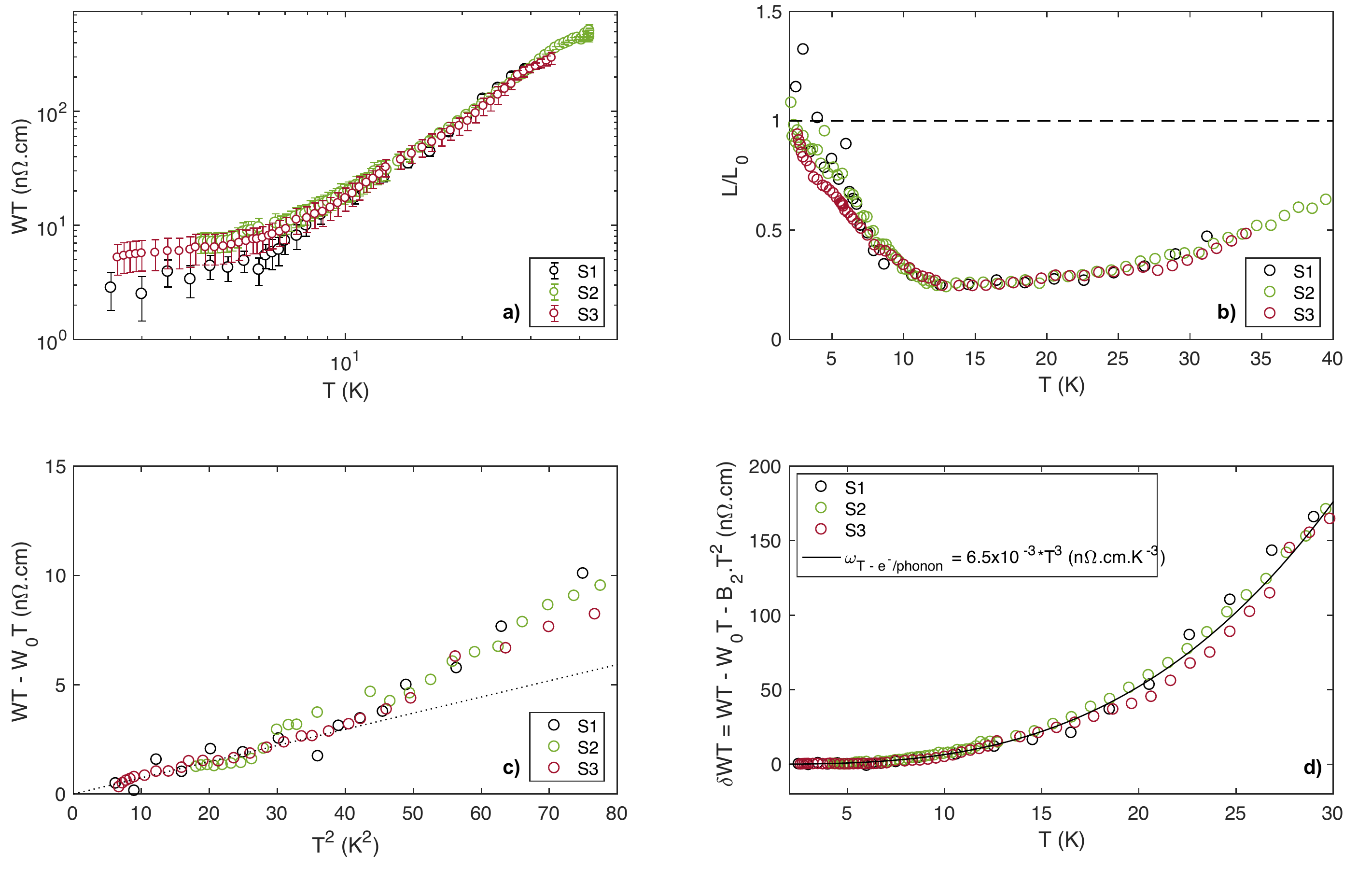}}
    \caption{a) Thermal resistivity, $WT=\frac{L_0T}{\kappa}$, as a function of $T$ for the three WP$_2$ samples. b) Ratio of Lorenz, $L(T)=\frac{\kappa}{T\sigma}$, to Sommerfeld, $L_0=2.44\times10^{-8}$ $W.\Omega.K^{-2}$, numbers as a function of temperature for the same three samples. c) Here we show $\,WT\,-\,W_0T$ as a function of $T^2$ for the three samples. d) Plot of the phonon component of the thermal resistivity $\delta WT\,=\,WT\,-\,W_0T\,-\,BT^2$ as a function of $T$. The black line corresponds to a $T^3$ fit.}
    \label{fig2}
\end{figure}
\end{center}

\section{Electrical and Thermal $T^2$-dependent resistivity : comparison of different systems}
We reference in table \ref{table2} the values of the electrical and thermal resistivities quadractic prefactors for different materials. The electrical prefactor is noted $A_2$ whereas the thermal $T^2$-prefactor is $B_2$. We computed the ratio of these two terms in the third column.
\begin{table}[h!]
\small
\begin{tabular}{|c||c|c|c|c|}
 \hline
 Material&Residual Resistivity $\rho_0$ (n$\Omega$.cm)& Electrical Prefactor $A_2$ (p$\Omega$.cm.K$^{-2}$) & Thermal Prefactor $B_2$ (p$\Omega$.cm.K$^{-2}$)& $A_2/B_2$ \\[0.5ex] 
 \hline\hline
 WP$_2$&4 - 7&16.6 &75.6 & 0.22 \\
 \hline
 W&0.06 - 0.5&0.9$\pm$0.3&6.2$\pm$0.9 &0.15 \\
 \hline
 UPt$_3$&200 - 600&(1.6$\pm$0.59)$\times$10$^6$ &(2.44$\pm$0.9)$\times$10$^6$ &0.65 \\
 \hline
 Ni&1 - 3&25$\pm$5 &61 &0.40 \\
 \hline
 CeRhIn$_5$&37&21000 &57000 &0.4 \\
 \hline
\end{tabular}

\caption{Presentation of the electrical ($A_2$) and thermal ($B_2$) quadratic prefactors of different materials. WP$_2$ data from this work, W from \cite{PhysRevB.3.3141}, UPt$_3$ from \cite{PhysRevLett.73.3294}, elemental Ni from \cite{nickel} and CeRhIn$_5$ from \cite{PhysRevLett.94.216602}.}
\label{table2}
\end{table}

\section{Quadratic Thermal Resistivity in low temperature liquid $^3$He}

\begin{figure}[H]
\makebox{\includegraphics[width=0.66\textwidth]
    {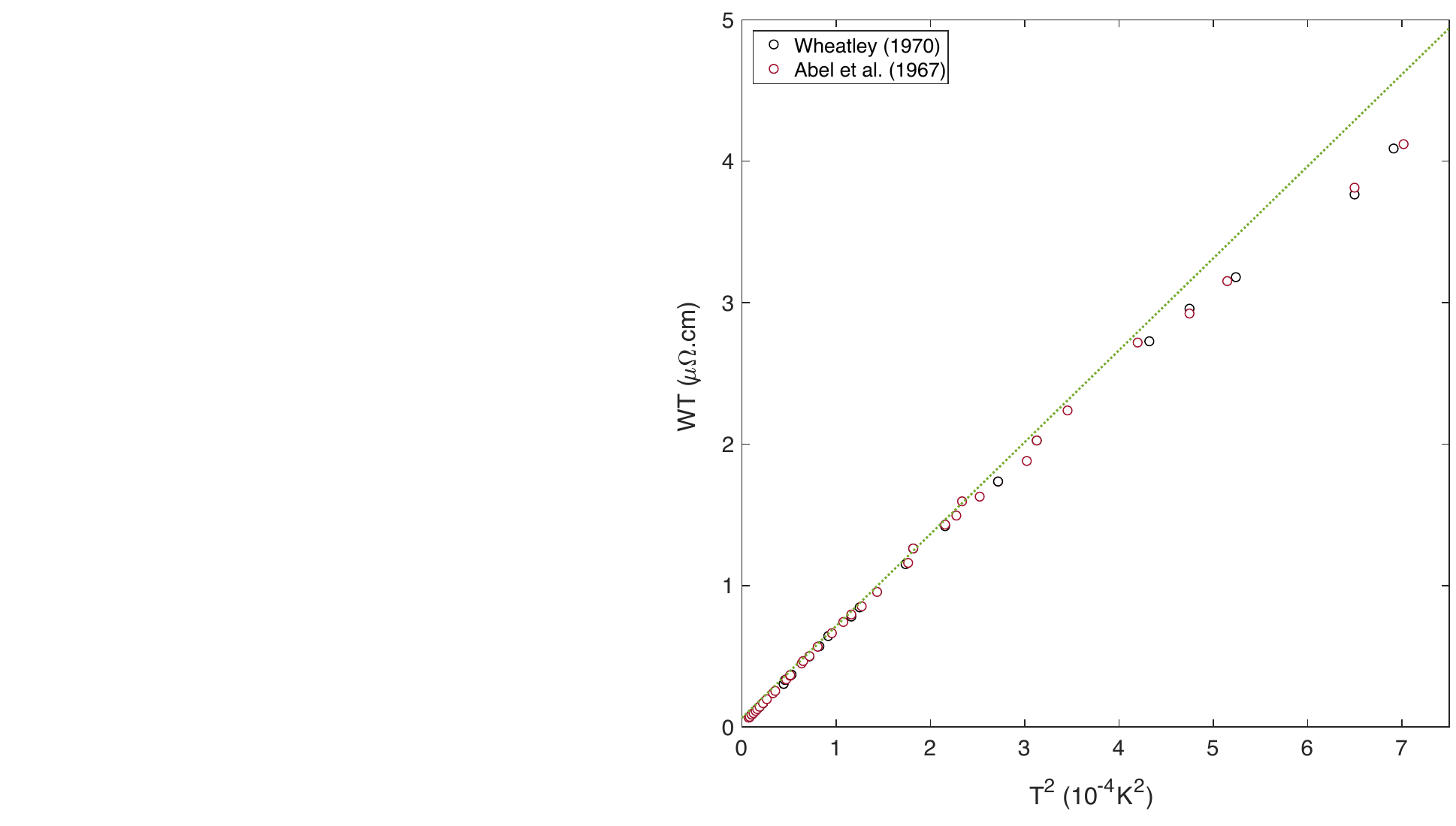}}
    \caption{Thermal resistivity $WT$ of saturated normal state liquid $^3$He from \cite{PhysRevLett.18.737,wheatley:jpa-00213853} as a function of $T^2$. WT is behaves as $T^2$ for temperatures between 2.6mK and 0.2K. The slope is the value used in the Kadowaki-Woods plot in the corpus.}
    \label{He3}
\end{figure}

The thermal conductivity, $\kappa$, of normal-liquid $^3$He was measured by Abel \emph{et al.} \cite{PhysRevLett.18.737} and Wheatley \cite{wheatley:jpa-00213853}.  They expressed their thermal conductivity as : $(\kappa T)^{-1}= a+bT$. At low temperature, the first term exceeds by far the second and $\kappa$ is basically proportional to the inverse of temperature, implying $WT \propto T^2$. This can be seen in figure \ref{He3}, where we plot $WT$ as a function of $T^2$.

\section{Specific Heat of WP$_2$}
\begin{figure}[H]
\makebox{\includegraphics[width=1.025\textwidth]
    {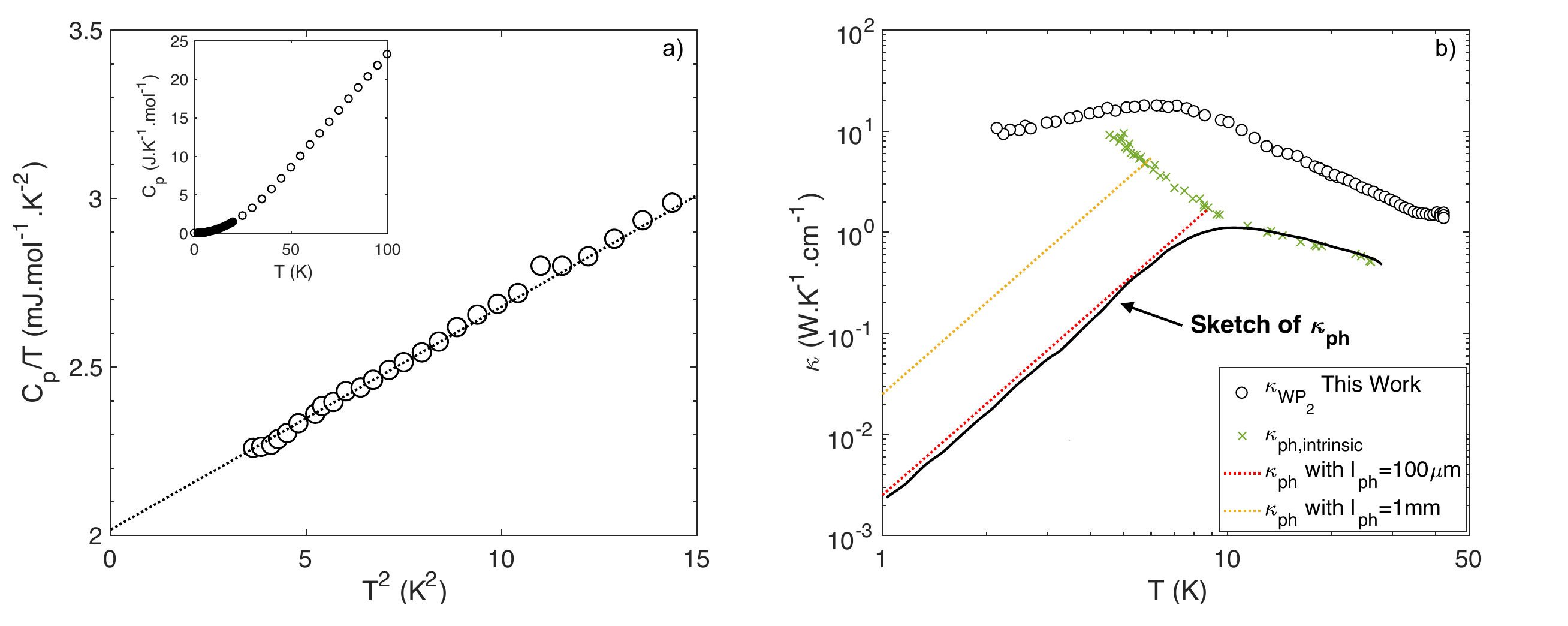}}
    \caption{a) Specific Heat plotted as $C_p/T$ as a function of $T^2$ for WP$_2$. The inset shows the specific heat plotted as $C_p$ as a function of $T$ up to 100K. b) Thermal conductivity, $\kappa$, as a function of temperature in WP$_2$ (black dots). The red dotted line is an upper limit to the phonons' contribution to thermal conductivity in WP$_2$ given the size of our crystal. whereas the yellow line corresponds to a mean free path of 1mm. Phonon thermal conductivity in Bi crystals of different sizes \cite{PhysRevLett.98.076603} are shown. The solid sketched line represents our estimation of maximum phonon contribution in WP$_2$.}
    \label{HC}
\end{figure}

In order to derive the specific heat due to phonons without the exact phonon dispersion curve we have to make an approximation. We will consider that for $T\ll\theta_D$ with $\theta_D=445K$, optical modes are not excited : the thermal energy is too small. We are left with acoustic modes only. This is the Debye approximation which yields a $T^3$-dependent phononic specific heat. But heat capacity can also result from electronic contribution. In that case it takes the form of linear in $T$ term. We define the specific heat from both contribution in equation \ref{eqHC} 
\begin{equation}
C_p=\gamma T + C_3.T^3
\label{eqHC}
\end{equation}
Fig.\ref{HC}.a presents the specific heat $C_p$ plotted as $C_p/T$ as a function of $T^2$ measured in our WP$_2$ samples.
Below 4K we can estimate both contributions. First, from the intercept with the \emph{y}-axis we determined the electronic contribution $C_{el}=\gamma T$, then the slope gave the phononic contribution $C_3=6.62\times 10^{-2}$ mJ.mol$^{-1}$.K$^{-4}$.
As temperature is furthered increased the $T^3$ contribution is suppressed from the specific heat and we head toward a saturating regime of $C_p$. This is observed in the inset of Fig.\ref{HC}.a. The Dulong-Petit law is then recovered.
Also, the heat capacity of phonons is linked to the thermal conductivity $\kappa_{ph}$ through the following equation :
\begin{equation}
\kappa_{ph}=\frac{1}{3}\,C_{ph}\times v_s\times l_{ph}
\label{eqkapandHC}
\end{equation}
Where $C_{ph}$ is the heat capacity per unit of volume, $v_s$ is the average speed of the phonons in the system and $l_{ph}$ the mean free path of the phonons. As a matter of fact, we can now derive an upper limit to an contribution to thermal conductivity caused by phonons in WP$_2$. To do so, we use the phonons' specific heat determined above, a speed of sound of $v_s=3000\,m.s^{-1}$ and a mean free path as high as possible in our sample, i.e 100$\mu$m. 

The phonons' contribution estimated for $l_{ph}=100\mu$m and $l_{ph}=1$mm are plotted respectively as a red dotted line and a yellow dotted line in Fig.\ref{HC}.b. In order to have an idea of the intrinsic, i.e size independent, contribution of the phonons to the thermal conductivity of the system, it is instructive to refer to the case of bismuth where the evolution of the thermal conductivity with size as been documented \cite{PhysRevB.23.449,PhysRevLett.98.076603}. This allows us to make a rough sketch of $\kappa_{phonon}$ in our system. As seen in the figure, the expected contribution of phonons remains an order of magnitude below the measured thermal conductivity in WP$_2$ in the whole temperature window.

We deduce from this observation that the phonons' contribution to $\kappa$ is negligible compared to the electronic contribution.
\vskip 1 \baselineskip
\newpage

\section{Residual resistivity decomposed to boundary and impurity scattering}
In the main text, we argued that Gurzhi's hydrodynamic criteria \cite{gurzhi} can be satisfied in a narrow temperature window. Here we show that the width of this window and its very existence is critically dependent on the importance of impurity scattering. Assuming a carrier density of $2.5\times10^{21}$cm$^{-3}$, the mean-free-path exceeds the sample width and the residual resistivity is entirely set by boundary scattering. This will lead to Fig.\ref{rho0}.a). In a limited temperature window (7$<$T$<$13 K) the MC, MR and boundary scattering times respect the hydrodynamic requirements.
Let us consider the possibility that the residual resistivity $\rho_0$ contains a sizable component due to impurity scattering. In this case, one can write:
\begin{equation}
\rho_0 = \rho_{00} + \rho_{imp}
\label{rho0}
\end{equation}
Here, $\rho_{00}$ is due to boundary scattering whereas $\rho_{imp}$ results of impurities and defects. When $\rho_{00}$ only represents 75\% (Fig\ref{rho0}.b)),  we can still find a temperature range to satisfy Gurzhi's conditions. This hydrodynamic window is shifted to lower temperatures and gets narrower. However, if $\rho_{00}$ becomes 50\% (Fig\ref{rho0}.c)) of the residual resistivity, then there is no range of temperature which allows the emergence of a hydrodynamic regime in WP$_2$.
\begin{figure}[H]
\makebox{\includegraphics[width=0.975\textwidth]
    {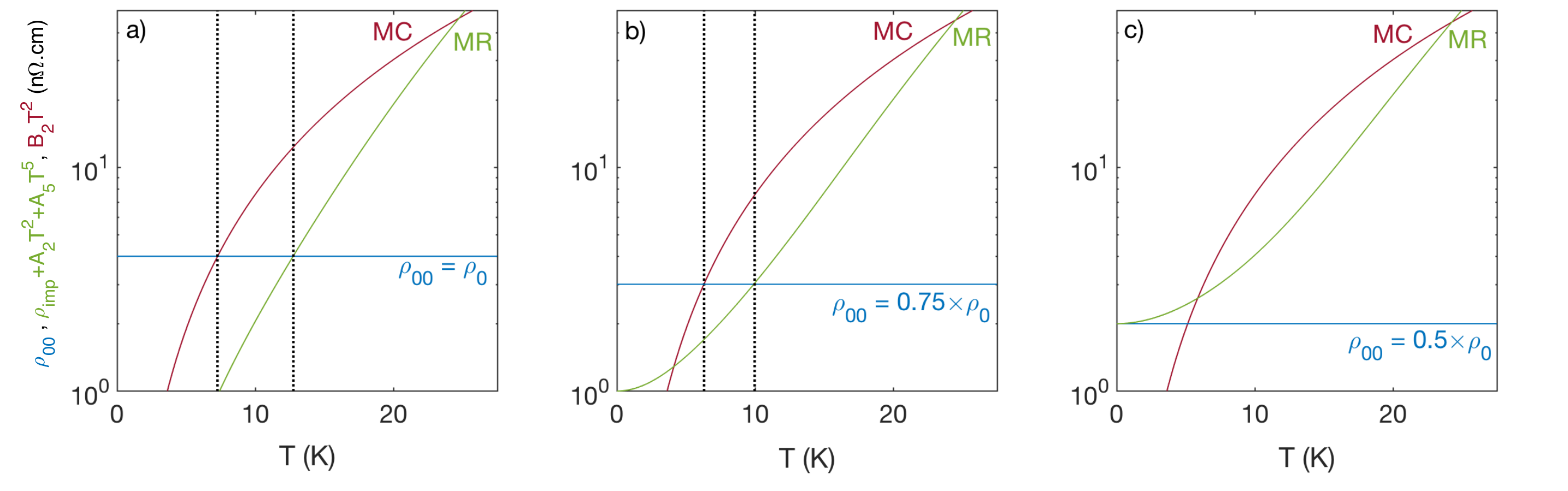}}
    \caption{The magnitude of B$_{2}T^{2}$, proportional to momentum-conserving (MC) electron-electron collisions is compared to A$_{2}T^{2}+$A$_{5}T^{5}$, which is proportional to momentum-relaxing (MR) collisions by electrons and phonons and $\rho_{00}$, which is a measure of boundary scattering. A limited window where the required hierarchy for hydrodynamics is satisfied can be found for the three following cases. a) $\rho_0$ is fully due to boundary scattering. b)$\rho_{00}$, due to boundary scattering, represents 75\% of the total residual resistivity. c) $\rho_{00}$, due to boundary scattering, represents 50\% of the total residual resistivity.}
    \label{regimes}
\end{figure}
\pagebreak
\section{Computational methods}

The density functional theory calculations were performed using the
general full-potential linearized augmented planewave method as
implemented in the {\sc wien2k} software package \cite{wien2k}. The
generalized gradient approximation of Perdew, Burke and Ernzerhof was
used for the exchange-correlation functional. The muffin-tin radii of
2.3 and 2.0 a.u.\ were used for W and P, respectively. A
$24\times24\times20$ $k$-point grid was used to perform the Brillouin
zone integration in the self-consistent calculations. The planewave
cutoff was set by $RK_{\textrm{max}}$ = 8, where $K_{\textrm{max}}$
is the planewave cutoff and $R$ is the smallest muffin-tin radius used
in the calculations.
The DFT calculations yields the following values :

- carrier concentration of holes due to band 1: 1.25$\times$ $10^{21}$ cm$^{-3}$ 

- carrier concentration of holes due to band 2: 1.65$\times$ $10^{21}$ cm$^{-3}$ 

- carrier concentration of electrons due to band 3: 1.69$\times$ $10^{21}$ cm$^{-3}$ 

- carrier concentration of electrons due to band 4: 1.18$\times$ $10^{21}$ cm$^{-3}$

\end{document}